\documentclass[12pt,a4paper]{article}
\usepackage{amssymb}
\usepackage{amsfonts}
\usepackage[english]{babel}
\usepackage{graphicx}
\usepackage{appendix}
\newcommand{\bean}{\begin{eqnarray*}}
\newcommand{\eean}{\end{eqnarray*}}
\newcommand{\ba}{\begin{array}}
\newcommand{\ea}{\end{array}}
\newcommand{\be}{\begin{equation}}
\newcommand{\ee}{\end{equation}}
\newcommand{\bea}{\begin{eqnarray}}
\newcommand{\eea}{\end{eqnarray}}
\newcommand{\pa}{\partial}

\newcommand{\no}{\nonumber}

\begin{document}

\title
{Asymptotic analysis of multi-lumps  solutions in the Kadomtsev-Petviashvili-(I) equation  }

\author{
 Jen-Hsu Chang \thanks{e-mail:jhchang@ndu.edu.tw} \\ Department of Computer Science and Information Engineering , \\
National Defense University, \\
 Tau-Yuan, Taiwan, 33551 }

\date{}

\maketitle
\begin{abstract}
Inspired by the works of Y. Ohta and J. Yang, one constructs the lumps solutions in the Kadomtsev-Petviashvili-(I) equation using the Grammian determinants. It is shown that the locations  of peaks  will depend on the real roots of Wronskian of the orthogonal polynomials for the asymptotic behaviors in some particular cases. Also, one can prove that all the locations of peaks are on a vertical line when time approaches - $\infty$, and then they will be on a horizontal line when time approaches $\infty$, i.e., there is a rotation $\frac{\pi}{2}$ after interaction.

\end{abstract}

\newpage

\section{Introduction }
The lumps solutions of Kadomtsev-Petviashvili(KP)-(I)equation correspond to the rationally decaying  solutions (like O ($\frac{1}{x^2+y^2}$)), and they were first found in \cite{mn} and constructed by direct algebraic methods \cite{ab3, ab4, kn}. It is known that the Schrodinger equation can be used to linearize the KP-(I) equation via the inverse scattering transformation ( see (\ref{li}) below). The lumps solutions of reflectionless potentials have discrete spectrum for the Schrodinger equation, and the associated wave functions are meromorphic functions with  respect to the spectrum; moreover, in general, the associated wave functions can be solved by the Fredholm-like integral equation \cite {ab4}. Also, the lumps solutions of DS equation were investigated in \cite{ms}. On the other hand, the asymptotic analysis of these lumps solutions are also noteworthy. These lumps can be thought as a collection of of individual humps that generically evolve with non-uniform dynamics, and their asymptotic behaviors, that is, $\vert t \vert \to \infty$, can be described by the locations and poles structure of the associated meromorphic wave functions in a complicated way \cite{ab3, ab4}. It is not clear how to describe the locations of the peaks of these lumps solutions as  $\vert t \vert \to \infty$ for KP-(I) equation. \\
\indent The basic shallow water waves equation of 2+1 integrable model is the KP equation describing small amplitude  with slow variations in the direction transverse to the wave propagation\cite{ch, ko, ko1, ko6} :
\be u_t+u_{xxx}+6uu_x \pm \pa_x^{-1}3u_{yy}=0,  \label{kp} \ee
where $u$ is the height of water wave and  $x,y$ are the space coordinates, t being the time. Here the subscripts denote partial derivatives and  
\[ \pa^{-1}_xu (x,y,t)= \int_{-\infty}^{x} u(x', y, t) dx' .\]
The "+" is the KP-(II) equation and the "-" is the KP-(I) equation. It depends on the surface tension of water waves. 
The integrability structure of the KP equation can be found in \cite{hi} and its physical origin,  the inverse scattering transformation  can be found in \cite{ab}.

To construct the lumps solutions of KP-(I) equation , one introduces the Grammian solution structure \cite{ab3, hi}. The KP-(I) equation can be written as the compatibility of the linear equations:
\bea 
&& i \psi_y +\psi_{xx}+u\psi =0 \no \\
&& \psi_t+4 \psi_{xxx}+6u\psi_x+ w\psi=0, \quad  w_x=u. \label{li}
\eea
The adjoint of (\ref{li}) is 
\bea 
&& -i \psi^*_y +\psi^*_{xx}+u\psi^* =0 \no \\
&& \psi^*_t+4 \psi^*_{xxx}+6(u\psi^*)_x- w\psi^*=0, \quad  w_x=u. \label{lic} \eea
Let $ \{\psi_j, \psi^*_j \} , j=1,2,3, \cdots, m $, be distinct solutions of (\ref{li}) and (\ref{lic}) respectively, corresponding to a given solution $u(x,y,t)$ of KP-(I) equation. Then a new solution $\hat u $ is given by \cite{ab3}
\be \hat u=u +2 \pa_x^2 \ln (det M(x,y,t)), \label{ru} \ee
where $det M (x,y,t)$ is the Grammian determinant of the $m \times m$ matrix with elements
\be M_{lj}=\int_{-\infty}^x \psi_l (x', y,t) \psi_j^* (x', y, t) dx', l, j=1,2,3, \cdots, m.  \label{it} \ee
\indent The paper is organized as follows. In section 2, we obtain
the  lumps solutions solutions using the Grammian determinant structure 
and the elementary Schur functions. Section 3 is used to analyze the asymptotic behaviors of lumps  solutions. It is shown that the locations of peaks  will depend on the real roots of Wronskian of the orthogonal polynomials for some special cases. The section 4 is devoted to the concluding remarks.

\section{Grammian Determinant Solutions} 
In this section,we construct the multi-lumps solutions  (\ref{ru}) of the KP-(I) equation.\\
\indent  Without loss of generality, we choose the seed solution $u=w=0$ for the asymptotic analysis. Inspired by the work in \cite{oy2, oy3}, then one considers the rogue waves solutions in KP-(I)equation. Let 
\bea 
\psi_l &= & A_l e^{\xi_l}, \quad \xi_l= p_l x+i p_l^2 y -4 p_l^3 t, \quad p_1, p_2, \cdots, p_m \in \textsl{C} \no \\
\psi_j^* &= & B_j e^{\eta_j}, \quad \eta_j= q_j x - iq_j y -4 q_j^3 t, \quad q_1, q_2, \cdots, q_m \in \textsl{C} \label{ro}
\eea
where $A_l$ and $B_j$ are differential operators defined by 
\be A_l= \sum_{k=0}^{n_l} c_{lk} (p_l\pa_{p_l})^{n_l-k}, \quad B_j= \sum_{s=0}^{n_j} d_{js} (q_j\pa_{q_j})^{n_j-s}, \label{de} \ee
where the coefficients $ c_{lk} $ and $d_{js}$ are complex numbers.  One supposes that 
\[ A_l e^{ \xi_l}= \textit{P}_l e^{ \xi_l}, \quad  B_j e^{ \eta}_j= \textit{Q}_j e^{ \eta_j},\]
where $\textit{P}_l(x,y,t)$ and $\textit{Q}_j(x,y,t)$ are functions of $x,y,t$. Then, using integration by parts, we can get
\be  M_{lj}=e^{\xi_l+\eta_j } \sum_{\nu=0}^{n_l+n_j} \frac{(-1)^{\nu}}{ (p_l+q_j)^{\nu+1}} \pa_x^{\nu} (\textit{P}_l \textit{Q}_j ). \label{in} \ee
One can take the following constraints to get real solutions of $u$
\be q_j=\bar{p}_j, \quad d_{lj}= \bar{c}_{lj}    \label{ra} \ee 
such that the matrix $M$ is Hermitian $M_{lj}= \bar{M}_{jl}$ and $ \textit{Q}_j= \bar{\textit{P}_j}$. \\

To find the functions $\textit{P}_l$, we define  $S_r(x,y,t,p)$ as follows: ($r \geq 0$ )
\be (p \pa_p)^r e^{\xi (p)} = S_r(x,y,t,p)e^{\xi (p)}, \quad \xi (p)=x p+i p^2 y -4 p^3 t. \label{sr} \ee
Then one has 
\bean   (p \pa_p)^{r+1}  e^{\xi (p)} &=& (p \pa_p) [(p \pa_p)^r e^{\xi (p)}]=(p \pa_p)[S_r(x,y,t,p)e^{\xi (p)}]  \\
  &=& p(\pa_p (S_r)+ f(x,y,t,p) S_r) e^{\xi (p)} = S_{r+1} e^{\xi (p)},
\eean
where 
\be  \frac{ \pa \xi (p)}{\pa p}=f(x,y,t,p)= x +2ipy-12p^2 t.\label{fu} \ee 
Hence we get 
\be S_{r+1}= p\pa_p (S_r)+ pfS_r.\label{gen} \ee
For example, 
\bean S_0 &=& 1, \\
S_1 &=& pf, \\
S_2 &=& pf+p^2f_p+p^2f^2 \\
S_3 &=& pf+3p^2 f^2+p^3 f^3+3p^2 f_p+ 3fp^3 f_p+p^3 f_{pp}, \\
S_4 &=& pf+ 6\,{p}^{3}f_{pp} +{p}^{4}f_{ppp}+18\,{p}^{3}f_p f +3\,{p}^{4}f_p^{2}+4\,{p}^{4}f f_{pp}+7\,{p}^{2}f_p +7\,{p}^{2}f^{2}+6\,{p}^{3}f^{3} \\
&+& 6\,{p}^{4}f^{2}f_p +{p}^{4} f^{4} 
\eean
Now, we have 
\[ \textit{P}_l= \sum_{k=0}^{n_l} c_{lk} S_{n_l-k}. \]
\indent On the other hand, we can express $S_r$ as elementary Schur functions in the following way. The elementary Schur functions $H_r$ are defined by 
\[ e^{\sum_{k=1}^{\infty} x_k \lambda^k } = \sum_{r=0}^{\infty}  H_r (\vec{x}) \lambda^r , \] 
where $\vec{x}=(x_1,x_2,x_3, \cdots)$.
In general, we have 
\be H_r (\vec{x})= \sum_{n_1+2n_2+3n_3+\cdots+rn_r=r} \frac{x_1^{n_1} x_2^{n_2}x_3^{n_3} \cdots x_r^{n_r}}{n_1 ! n_2 ! \cdots n_r !}.  \label{sc} \ee 
For example, 
\bean && H_0=1, \quad H_1=x_1, \quad  H_2=\frac{1}{2} x_1^2+x_2, \quad H_3=\frac{1}{6} x_1^3+x_1x_2+ x_3, \no \\
&& H_4=x_4+ x_1 x_3+\frac{1}{2}x_2^{2}+\frac{1}{2} x_2 x_1^{2}+\frac{1}{24}x_1^{4} \no \\
&& H_5= x_5+x_1x_4+x_3x_2+\frac{1}{2}x_3x_1^2+\frac{1}{2}x_2^2x_1+\frac{1}{6}x_2x_1^3+\frac{1}{120}x_1^5, \cdots.  \eean
We remark that $ H_r $ has the basic property 
\be  \frac{\pa H_r}{\pa x_s}=H_{r-s}. \label{ba} \ee
Using the functional identity, 
\[ e^{\lambda p\pa_p } F(p)= F(e^{\lambda} p), \]
one has,  defining  that $\xi=xp+ip^2y-4p^3t=px_1+p^2 x_2+p^3x_3=\sum_{\mu=1}^3 p^{\mu}x_{\mu}$   ,
\bea
e^{-\xi } e^{\lambda p\pa_p } e^{\xi } &=& \exp(\sum_{\mu=1}^3 (e^{\mu \lambda}-1)p^{\mu}x_{\mu})=\exp(\sum_{r=1}\frac{\lambda^{r}}{r!} \sum_{\mu=1}^3 \mu^{r}p^{\mu} x_{\mu}) \label{sh} \\
&=& \sum_{r=0}^{\infty} \lambda^r H_r (\vec{\chi}_r(p)), \no
\eea
where $ \vec{\chi}_r(p)=(\chi_1(p), \chi_2 (p), \chi_3(p), \cdots, \chi_r(p) )= (x_1, x_2, x_3, \cdots, x_r) $, i.e.,  
\be x_n= \chi_n(p)= \frac{\sum_{\mu=1}^3 \mu^n p^{\mu} x_{\mu} }{n!}=\frac{ xp +i p^2 2^{n} y -4p^3 3^n t }{n!}, \quad n=1, 2, 3, \cdots, r .\label{va} \ee
Comparing the coefficient of $\lambda^r$ of (\ref{sh}), one gets 
\[ e^{-\xi } \frac{(p\pa_p)^r}{r!}  e^{\xi } = H_r(\vec{\chi}_r(p)).\] 
Hence 
\be  S_r= r! H_r(\vec{\chi}_r(p)).\label{eq} \ee
For the case $m=1$ in (\ref{ru}), we obtain  $u=2 \pa_{xx}^2 \ln F_n,$ where 
\be F_n=M_{11}= \sum_{\nu=0}^{n} \frac{(-1)^{\nu}}{ (p_1+\bar{p}_1)^{\nu+1}} \pa_x^{\nu} (\vert \textit{P}_1 \vert^2 ), \label{pos} \ee
and
\[ \textit{P}_1= c_{10} S_n+ c_{11} S_{n-1}+c_{12} S_{n-2}+ \cdots+ c_{1n} S_0.\] 
\indent Under the condition (\ref{ra}), the matrix (\ref{it}) is Hermitian and positive. It can be seen as follows. For any non-zero column vector $\textbf{v}=(v_1, v_2, \cdots, v_m)^T$ and $\bar{\textbf{v}}$ being its complex transpose, we have 
\bea \bar{\textbf{v}} M \textbf{v}&=& \sum_{l, j=1}^m \bar{v}_l M_{lj}v_j = \sum_{l, j=1}^m \bar{v}_l v_j A_l B_j \int_{-\infty}^x e^{\xi_l +\eta_j} dx  \no \\
&=& \int_{-\infty}^x (\sum_{l, j=1}^m \bar{v}_l v_j A_l B_j  e^{\xi_l +\eta_j})  dx  =\int_{-\infty}^x \vert  \sum_{l=1}^m \bar{v}_l  A_l  e^{\xi_l} \vert^2 dx. \label{her}
\eea 
\indent  From (\ref{it}) it follows that 
\be M_{lj}= A_l B_j \frac{1}{p_i+q_j} e^{\xi_l +\eta_j}. \label{ter} \ee
By using the operator relation 
\[ (p_l\pa_{p_l}) e^{\xi_l} =e^{\xi_l}(p_l\pa_{p_l} + \hat{\xi}_l), \quad (q_j\pa_{q_j}) e^{\eta_j} =e^{\eta_j }(q_j\pa_{q_j} + \hat{\eta}_j), \]
where
\bean
\hat{\xi_l} &=& p_l \pa_{p_l} \xi_l =xp_l+2ip_l^2y-12p_l^3t, \\
\hat{\eta}_j &=& q_j \pa_{q_j} \eta_j =xq_j-2iq_j^2y-12q_j^3t,
\eean
we also have 
\be M_{lj}= e^{\xi_l+\eta_j }\sum_{k=0}^{n_l} c_{lk}(p_l\pa_{p_l} + \hat{\xi_l})^{n_l-k } \sum_{s=0}^{n_j} d_{js}(q_j\pa_{q_j} + \hat{\eta}_j)^{n_j-s} \frac{1}{p_l+q_j}.\label{im} \ee
\section{Asymptotic Analysis} 
In this section, we can investigate the asymptotic behaviors of the multi-lumps solutions  defined by (\ref{it}) , (\ref{in}) and (\ref{ra}) or (\ref{im}). From this asymptotic analysis, one can get the orthogonal polynomials, of which the real roots are used to determine the locations of peaks as $\vert t \vert \to \infty$.  Without loss of generality, we assume that $ c_{10}=c_{20}=\cdots =c_{n0}=1$ by (\ref{ru}).  Let's define 
\[ \hat {A}_j= \sum_{k=0}^{n_j} c_{jk}(p_j\pa_{p_j} + \hat{\xi_j})^{n_j-k }, \]
and it's not difficult to  see that 
\be [\hat {A_l}, \hat {A_j}]=0, \quad [\hat {A_l}, \bar{\hat {A}_j}]=0, \quad [ \bar{\hat {A_l}}, \bar{\hat {A_j}}]=0, \quad l \neq j. \label{co} \ee
Notice that the differential operators in (\ref{de}) also have the properties (\ref{co}). By (\ref{co}) and the reality condition (\ref{re}), one has the expression 
\bea \tau(x,y,t) &:=& e^{-(\sum_{j=1}^{m} p_j+ \bar{p}_j)} det M(x,y,t) \no \\
&=& e^{-(\sum_{j=1}^{m} p_j+ \bar{p}_j)} A_1 A_2 \cdots A_m \bar{A_1}\bar{A_2} \cdots  \bar{A_m} \no \\
&& \frac{\prod_{1 \leq l < j \leq m} (p_l-p_j)( \bar{p}_l-\bar{p}_j)}{\prod_{l,j=1}^{m} (p_l+\bar{p}_j)} e^{\sum_{j=1}^{m} p_j+ \bar{p}_j} \label{cb} \\
&=&\hat {A_1} \hat {A_2} \cdots \hat {A_m} \bar{\hat {A_1}}\bar{\hat {A_2}} \cdots \bar{\hat {A_m}} \frac{\prod_{1 \leq l < j \leq m} (p_l-p_j)( \bar{p}_l-\bar{p}_j)}{\prod_{l,j=1}^{m} (p_l+\bar{p}_j)}, \label{ca} \eea
where one uses the Cauchy determinant formula 
\[ det (\frac{1}{ x_i +y_j})_{1 \leq i, j \leq m}=\frac{\prod_{1 \leq i < j \leq m} (x_i-x_j)( y_i-y_j)}{\prod_{i,j=1}^{m} (x_i+y_j)}. \]
For example, when $n=3$, we have 
\bean && \tau(x,y,t)= det  \left(\ba{ccc}   \hat {A_1} \bar{\hat {A_1}} \frac{1}{ p_1 +\bar{p}_1}   &  \hat {A_1} \bar{\hat {A_2}} \frac{1}{ p_1 +\bar{p}_2}  & \hat {A_1} \bar{\hat {A_3}} \frac{1}{ p_1 +\bar{p}_3} \\
\hat {A_2} \bar{\hat {A_1}} \frac{1}{ p_2 +\bar{p}_1}  & \hat {A_2} \bar{\hat {A_2}} \frac{1}{ p_2 +\bar{p}_2}  & \hat {A_2} \bar{\hat {A_3}} \frac{1}{ p_2 +\bar{p}_3} \\
\hat {A_3} \bar{\hat {A_1}} \frac{1}{ p_3 +\bar{p}_1} & \hat {A_3} \bar{\hat {A_2}} \frac{1}{ p_3 +\bar{p}_2}  & \hat {A_3} \bar{\hat {A_3}} \frac{1}{ p_3 +\bar{p}_3}  \ea \right) \\ 
&=&  \hat {A_1} \hat {A_2}  \hat {A_3} \bar{\hat {A_1}}\bar{\hat {A_2}}\bar{\hat {A_3}} det 
 \left(\ba{ccc} \frac{1}{ p_1 +\bar{p}_1}  & \frac{1}{ p_1 +\bar{p}_2}  & \frac{1}{ p_1 +\bar{p}_3} \\
                \frac{1}{ p_2 +\bar{p}_1}  & \frac{1}{ p_2 +\bar{p}_2}  & \frac{1}{ p_2 +\bar{p}_3}  \\
                \frac{1}{ p_3 +\bar{p}_1}  & \frac{1}{ p_3 +\bar{p}_2}  & \frac{1}{ p_3 +\bar{p}_3}  \ea \right) \\
&=&  \hat {A_1} \hat {A_2}  \hat {A_3} \bar{\hat {A_1}}\bar{\hat {A_2}}\bar{\hat {A_3}}   \frac{(p_1-p_2)(p_1-p_3)(p_2-p_3)(\bar{p}_1-\bar{p}_2)(\bar{p}_1-\bar{p}_3)(\bar{p}_2-\bar{p}_3)}{\prod_{l,j=1}^{3} (p_l+\bar{p}_j)}. \eean 
From (\ref{ru}), (\ref{in}) and (\ref{eq}), we see that the solution $u(x,y,t)=2 \pa_x^2 \ln \tau (x,y,t) $ can also be expressed in terms of the elementary Schur functions (\ref{sc}). As $\sqrt{x^2+y^2} \to \infty $, the leading order behavior of $ \tau(x,y,t) $ consists of terms arising from where all the highest-order differential operators, $(p_1 \pa{p_1})^{n_1} , (p_2\pa{p_2})^{n_2}, \cdots, (p_m\pa{p_m})^{n_m}$ and their complex conjugates, act on the factor $ \prod_{1 \leq l < j \leq m} (p_l-p_j)( \bar{p}_l-\bar{p}_j)  e^{\sum_{j=1}^{m} p_j+ \bar{p}_j} $ only. It is similar to the case in \cite{ab3}. This yields 
\[ \tau \approx W \bar{W}, \] 
where 
\bea W &=& e^{-\sum_{j=1}^{m} p_j} (p_1 \pa{p_1})^{n_1}(p_2\pa{p_2})^{n_2}\cdots(p_m\pa{p_m})^{n_m} \prod_{1 \leq l < j \leq m} (p_l-p_j) e^{\sum_{j=1}^{m} p_j} \no \\
       &=& Wr(S_{n_1}(p_1),S_{n_2}(p_2), S_{n_3}(p_3), \cdots, S_{n_m}(p_m)) = det \pa_x^{m-1}(S_{n_j}(p_j)) \label{wo} \eea
is the Wronskian of the elementary Schur polynomials of degrees $n_1, n_2, n_3, \cdots, n_m$. Here we have used (\ref{sr}) and the Vandermonde determinant formula. Furthermore, using (\ref{ba}), (\ref{va}) and (\ref{eq}), we can get 
\bean  \frac{\pa S_{n_j}}{\pa x} &=&  (n_j) ! (\frac{\pa H_{n_j}}{\pa x_1}\frac{\pa x_1}{\pa x}+ \frac{\pa H_{n_j}}{\pa x_2}\frac{\pa x_2}{\pa x}+ \cdots + \frac{\pa H_{n_j}}{\pa x_{n_j}}\frac{\pa x_{n_j}}{\pa x})   \\ 
&=& (n_j)! p (H_{n_j-1}+ \frac{1}{2!}H_{n_j-2} + \frac{1}{3!}H_{n_j-3}+ \cdots + \frac{1}{(n_j)!}H_0).\eean 
Finally, one yields, up to an overall constant factor $ p^{\frac{m(m-1)}{2}}(n_1)!(n_2)! \cdots (n_m)! $, 
\be  \tau \approx W \bar{W} \approx \Omega \bar{\Omega}, \quad as \quad \sqrt{x^2+y^2} \to \infty  \label{as} \ee
where \bean
 \Omega(x,y,t) = det 
 \left(\ba{ccccc} H_{n_1}(p_1)  & H_{n_2}(p_2) & H_{n_3}(p_3) &\cdots  & H_{n_m}(p_m)\\
                 H_{n_1-1}(p_1) &H_{n_2-1}(p_2)  & H_{n_3-1}(p_3) & \cdots  &H_{n_m-1}(p_m) \\
                H_{n_1-2}(p_1)& H_{n_2-2}(p_2) & H_{n_3-2}(p_3)&\cdots  & H_{n_m-2}(p_m) \\
                \vdots & \vdots & \vdots & \vdots & \vdots \\
                H_{n_1-(m-1)}(p_1) & H_{n_2-(m-1)} (p_2)&  H_{n_3-(m-1)}(p_3)& \cdots &  H_{n_m-(m-1)} (p_m)\ea \right).\eean
We see that  $ \tau (x,y,t)= O (\vert f \vert^{2\rho})$, $\rho=\sum_{j=1}^{m}n_j- \frac{m(m-1)}{2}$.  \\
\indent Next, we assume $p_1=p_2=\cdots= p_m=a+bi$. In the limit  $t \to \pm \infty$,  $u (x,y,t) $  generically has $\rho$ distinct peaks whose locations are asymptotically given by \cite{ab3}
\bea 
x^{(j)}(t) &=& 12 (a^2+b^2)t + c_j \vert t \vert^q+o(\vert t \vert^q) \no \\
y^{(j)}(t) &=& 12 a t + w_j \vert t \vert^q+o(\vert t \vert^q), \quad j=1,2,3, \cdots, \rho, \quad   1/3  \leq  q \leq 1/2, \label{asy}      \eea 
where $q$ depends on $m$ and $n_1, n_2, n_3, \cdots, n_m$. We remark here why $ 1/3  \leq  q \leq 1/2 $ is not explained in \cite{ab3}. The real constants $c_j$ and $w_j$ can be found as follows. We plug the forms (\ref{asy}) into (\ref{as}) and then we have 
\be \Omega(x,y,t)=(M(c, w; a,b)+ R(c,w; a,b) i ) \vert t \vert^{\rho q}+o(\vert t \vert^{\rho q}), \label{rea}\ee
where $ M(c,w)$ and $ R(c,w) $ are real polynomials of $c$ and $w$. Letting $ M(c,w; a,b)= R(c,w; a,b)=0$, we can get exactly $\rho$ (topological charge) real roots $(c_j, w_j)$, counting multiplicity, that is, 
\be M(c_j, w_j; a,b)= R(c_j, w_j; a,b)=0. \label{com} \ee
 Also, from (\ref{fu}) and (\ref{asy}), one has 
\be f=(c-2bw+2i aw ) \vert t \vert^q + o(\vert t \vert^q), \quad f_p= -24at+2i w \vert t \vert^q, \quad f_{pp}=-24t, \quad f_{ppp}=\cdots=0 \label{de2}. \ee
To find $q$, we define the weights for $S_n$ in  (\ref{sc}) by
\be  weight (f) =1, \quad weight (f_p) =2,  \quad weight (f_{pp}) =3, \quad weight ( f_{ppp})=4,   \cdots.  \label{deg} \ee
In the expansion (\ref{as}), to obtain the highest powers in $\vert t \vert $ by (\ref{eq}) and (\ref{de2}), we consider only the terms $ f^{\alpha} f_p^{\beta} f_{pp}^{\gamma} $ ( $\alpha, \beta, \gamma \geq 0 $) of the  highest weight $\rho$. Then we have 
\[ \alpha + 2 \beta+ 3 \gamma= \rho, \quad   \alpha q+ \beta+ \gamma=\rho q. \]
The right-hand equation means we assume these  terms have the highest power in $\vert t \vert $. So 
\be  q=\frac{\beta+ \gamma}{ \rho-\alpha}= \frac{\beta+ \gamma}{2 \beta+ 3 \gamma }. \label{qd} \ee
It is not difficult to see that $  1/3  \leq  q \leq 1/2 $.  Also, let's find the condition that the highest power in $\vert t \vert $ is the same. Suppose 
$ f^{\alpha^{'} } f_p^{\beta^{'} }f_{pp}^{\gamma^{'}} $ is another term having the highest  power in $\vert t \vert $, i.e.,  from (\ref{qd}), 
\[ \frac{\beta+ \gamma}{2 \beta+ 3 \gamma }=\frac{\beta^{'}+ \gamma^{'}}{2 \beta^{' } + 3 \gamma^{'}}.  \] 
Then  a simple calculation  gets 
\be \beta \gamma^{'}= \gamma \beta^{'}, \quad or \quad  \frac{\beta}{\beta^{'}}=\frac{ \gamma}{ \gamma^{'}}. \label{con} \ee
The equation (\ref{con}) will  imply the condition of those  terms having the highest  power in $\vert t \vert $. \\
\indent Now, we observe that each elementary Schur polynomial $H_r$ in (\ref{sc}) has the highest degree $r$ in $x_1= xp+2ip^2y-12p^3t$. One utilizes the Maple software to do the following computations. For simplicity, we set $a=1, b=0$, i.e, $p=1$. By (\ref{va}) and (\ref{asy}) one has 
\bea x_1 &=& x+2iy-12t=(c+2iw ) \vert t \vert^q + o(\vert t \vert^q), \no \\
 x_2 &= & \frac{x}{2}+2yi-18t= -12t + (\frac{c}{2}+2iw ) \vert t \vert^q + o(\vert t \vert^q). \label{pol} \eea 

\subsection{ $m=1, n_1=n=\rho $}
Among the terms of highest weight in  $S_n$, from (\ref{de2}) the terms $f^n$ and $ f^{n-2} f_p $ of weight n has the highest powers in $t$, that is, $\alpha=n-2, \beta=1, \gamma=0$ in (\ref{qd}).  Then $nq=(n-2)q+1.$ Hence  $q=1/2$. For example, in $S_4$, the terms $f_{ppp}, f_p^2, ff_{pp}, f^2 f_p, f^4$ have  the highest weight 4. We see that $f^4$ and $f^2 f_p$ has the highest powers $t^{4q}$ and $t^{2q+1}$ from (\ref{de2}). Then we get $q=1/2$. 
\begin{itemize}
\item For $ t \to -\infty$, we consider $c=0$ in (\ref{pol}), and as $t \to -\infty$ , one  gets $ \sqrt{x^2+y^2} \to \infty $ and consequently, the asymptotic expansion (\ref{as}) is used. Plugging (\ref{pol}) into $H_r$, we assume 
\be H_r(t)= \frac{\Gamma_r (w)}{ r !}  (2i)^r (-t)^{r/2}+ O (\vert t \vert ^{\frac{r-1}{2}}), \label{ga} \ee
where $ \Gamma_r(w)$ is a polynomial in $w$. To find the polynomials $\Gamma_r(w)$, we see that from (\ref{ga})
\bea  \frac{\pa H_r}{\pa t} &=&  \frac{\Gamma_r (w)}{ r !}  (2i)^r \frac{r}{2} (-t)^{(r-2)/2}(-1)+ O (\vert t \vert ^{\frac{r-3}{2}})  \no \\
&=& (\frac{\pa H_r}{\pa x_1}\frac{\pa x_1}{\pa t}+ \frac{\pa H_r}{\pa x_2}\frac{\pa x_2}{\pa t}+ \cdots + \frac{\pa H_r}{\pa x_r}\frac{\pa x_r}{\pa t})  \no  \\ 
&=&  - \frac{ \Gamma_{r-1}(w)}{(r-1)!}(2i)^{r-1 }2i w (-t)^{(r-1)/2} \frac{1}{2\sqrt{-t}}+ (-12 ) \frac{\Gamma_{r-2}(w)}{(r-2)!}(2i)^{r-2}(-t)^{(r-2)/2} \no \\
& + & O(\vert t \vert^{\frac{r-3}{2}}).\label{ca2} \eea 
Then one  has the recursive relation after a simple calculation,   
\be w \Gamma_r (w)=  \Gamma _{r+1}(w)+6 r\Gamma _{r-1}(w), \quad \Gamma_0=1, \quad \Gamma_1=w. \label{rec} \ee
Also, it follows 
\be  \frac{d \Gamma_r(w)}{d w}=  r \Gamma_{r-1}(w) \label{re} \ee
from 
\bea  \frac{\pa H_r}{\pa w } &=& (\frac{\pa H_r}{\pa x_1}\frac{\pa x_1}{\pa w}+ \frac{\pa H_r}{\pa x_2}\frac{\pa x_2}{\pa w}+ \cdots + \frac{\pa H_r}{\pa x_r}\frac{\pa x_r}{\pa w})  \no  \\ 
&=&  (2i)^{r-1} (-t)^{(r-1)/2} \frac{\Gamma_{r-1}(w)}{(r-1)!}  2i \sqrt{-t} + O (\vert t \vert^{\frac{r-2}{2}}) .\label{ca}\eea 
By (\ref{rec}) and (\ref{re}), it can be seen that $\Gamma_r (w)$ satisfies the linear second-order  differential equation  ( $ r \geq 0 $) 
\be \frac{d^2 \Gamma_r(w)}{d w^2} -\frac{w}{6}\frac{d \Gamma_r(w)}{d w}+  \frac{r}{6} \Gamma_r (w)=0. \label{he} \ee
The orthogonal polynomial solutions (\ref{rec}) of (\ref{he}) have r real roots for $\Gamma_r(w)$, as is proved in page 22 \cite{is} using the Christoffel-Darboux identity.  We list a few polynomials for reference. For example, 
\bean {\Gamma}_0(w) &=&1, \quad  {\Gamma}_1(w)=w,   \quad {\Gamma}_2(w)=(w^2-6), \\
 {\Gamma}_3(w)&=& w^3-18w, \quad {\Gamma}_4(w)=w^4-36 w^2+108, \quad {\Gamma}_5(w)=w^5-60 w^3+540w,  \\
{\Gamma}_6(w) &=& w^6-90 w^4+1620w^2-3240,  \quad {\Gamma}_7(w)=w^7-126 w^5+3780w^3-22680w, \\
{\Gamma}_8(w)& =&w^8-168 w^6+7560w^4-90720w^2+136080.    \eean
We remark that the terms in $\Gamma_r(w)$ come from the condition (\ref{con}). 
\item For $ t \to \infty$, the calculation is similar. But we consider $w=0$ in (\ref{pol}). Plugging (\ref{pol}) into $H_r$, we assume 
\be H_r(t)= \frac{G_r (c)}{ r !}  (t)^{r/2}+ O (\vert t \vert ^{\frac{(r-1)}{2}}), \label{ga2} \ee
where $ G_r(c)$ is a polynomial in $c$. To find the polynomials $G_r(c)$, in a similar way as (\ref{ca2}) and (\ref{ca}), we have 
\bea  c G_r (c) &=&   G_{r+1}(c)+24 r G_{r-1}(c), \quad G_0=1, \quad G_1=c, \label{rec3}  \\
 \frac{d G_r(c)}{d c} &= & r G_{r-1}(c).  \label{rec2} \eea
By (\ref{rec3}) and (\ref{rec2}), it can be also seen that $G_r (c)$ satisfies the linear second-order  differential equation  ( $ r \geq 0 $) 
\be \frac{d^2 G_r(c)}{d c^2} -\frac{c}{24}\frac{d G_r(c)}{d c}+  \frac{r}{24} G_r (c)=0. \label{he2} \ee
Also, the orthogonal polynomial solutions (\ref{rec3}) of (\ref{he2}) have r real roots for $G_r(c)$.  We list a few polynomials for reference.
For example, 
\bean G_0(c) &=&1, \quad  G_1(c)=c,   \quad G_2(c)=(c^2-24), \\
 G_3(c)&=& c^3-72c, \quad G_4(c)=c^4-144 c^2+1728, \quad G_5(c)=c^5-240 c^3+8640c,  \\
G_6(c) &=& c^6-360 c^4+25920c^2-207360,  \quad G_7(c)=c^7-504c^5+60480c^3-1451520c, \\
G_8(c)& =& c^8-672 c^6+120960c^4-5806080c^2+34836480. \eean
\end{itemize}
\indent Therefore, from (\ref{asy})one knows that  all the peaks  are  on  one vertical line when time approaches - $\infty$, and then they will be on a horizontal line when time approaches $\infty$. Consequently, there is a rotation $\frac{\pi}{2}$ after interaction of lumps.
\subsection{ $m=2, n_1=1, n_2=n, \rho=n $}
In this case, by (\ref{as}), we have 
\be  
 \Omega_{(1,n)}(x,y,t) = det 
 \left(\ba{cc} H_{1}(1)  & H_{n}(1) \\
                 1 &H_{n-1}(1) \ea \right)= H_{1}(1)H_{n-1}(1)- H_{n}(1)   \label{tw} \ee
To find $q$, we consider it in the following way. One focus on the terms $ f^n $ and $f^{n-2} f_p$ in $S_n$, that is, $x_1^n$ and $x_1^{n-2} x_2$ in $H_n$. By (\ref{sc}), the coefficient of $x_1^{n-2} x_2$ in $H_n$ is $\frac{1}{(n-2)!}$; consequently, the coefficient of this term in (\ref{tw}) is $ \frac{1}{(n-3)!}-\frac{1}{(n-2)!}$. Hence when $n=3$, we get $q=1/3$; in other cases, we get $nq=(n-2)q+1$, i.e, $q=1/2$. \\
\indent When $n=3$ and $ t \to -\infty$, one has in (\ref{rea})
\[  M(c,w)=  1/3\,{c}^{3}-4\,c{w}^{2}-16, \quad R(c,w)=2 {c}^{2}w-8/3\,{w}^{3};\]
and $ t \to \infty$, one has 
\[M(c,w)=  1/3\,{c}^{3}-4\,c{w}^{2}+16, \quad R(c,w)=2 {c}^{2}w-8/3\,{w}^{3}.\]
A  little algebra shows both of them have three real roots. So there are three peaks when $\vert t \vert \to \infty $. Also, after interaction, the peak in the $x$-axis, that is $w=0$, will slow down.  \\
\indent On the other hand, from (\ref{pol}), the real roots in (\ref{com}) for $q=1/2$  come from either $c=0$ or $w=0$  as $ t \to \pm \infty $(see below). Therefore, there are two more cases to be discussed. 
\begin{itemize}
\item As $t \to -\infty$, we consider $w=0$ in (\ref{pol}). Then $x_1=c  \sqrt{-t}$ and $x_2=-12t+\frac{c}{2} \sqrt{-t}$. Plugging (\ref{pol}) into $H_r$, we assume 
\be H_r(t)= \frac{\hat \Gamma_r (c)}{ r !}  (-t)^{r/2}+ O (\vert t \vert ^{\frac{r-1}{2}}), \label{cga} \ee
where $\hat  \Gamma_r(w)$ is also a polynomial in $c$. To find the polynomials $\hat \Gamma_r(c)$, we see that from (\ref{cga})
\bea  \frac{\pa H_r}{\pa t} &=&  \frac{\hat \Gamma_r (c)}{ r !}  \frac{r}{2} (-t)^{(r-2)/2}(-1)+ O (\vert t \vert ^{\frac{r-3}{2}})  \no \\
&=& (\frac{\pa H_r}{\pa x_1}\frac{\pa x_1}{\pa t}+ \frac{\pa H_r}{\pa x_2}\frac{\pa x_2}{\pa t}+ \cdots + \frac{\pa H_r}{\pa x_r}\frac{\pa x_r}{\pa t})  \no  \\ 
&=&  - \frac{ \hat \Gamma_{r-1}(c)}{(r-1)!}  (-t)^{(r-1)/2} \frac{c}{2\sqrt{-t}}+ (-12 ) \frac{ \hat \Gamma_{r-2}(c)}{(r-2)!}(-t)^{(r-2)/2} \no \\
& + & O (\vert t \vert^{\frac{r-3}{2}}).\label{cca} \eea 
Then one has the recursive relation after a simple calculation,   
\be c \hat \Gamma_r (c)=  \hat \Gamma _{r+1}(c)-24 r\hat \Gamma _{r-1}(c), \quad \Gamma_0=1, \quad \Gamma_1=c. \label{rec} \ee
When compared with the recursive relation (\ref{rec3}), the equation (\ref{rec}) has a minus sign before 24. 
Also, it follows 
\be  \frac{d \hat \Gamma_r(c)}{d c}=  r \hat \Gamma_{r-1}(c) \label{cre} \ee
from 
\bea  \frac{\pa H_r}{\pa c } &=& (\frac{\pa H_r}{\pa x_1}\frac{\pa x_1}{\pa w}+ \frac{\pa H_r}{\pa x_2}\frac{\pa x_2}{\pa w}+ \cdots + \frac{\pa H_r}{\pa x_r}\frac{\pa x_r}{\pa w})  \no  \\ 
&=&  (-t)^{(r-1)/2} \frac{\hat \Gamma_{r-1}(c)}{(r-1)!} \sqrt{-t} + O (\vert t \vert^{\frac{r-1}{2}}) .\label{cca2}\eea 
By (\ref{rec}) and (\ref{cre}), it can be seen that $\hat \Gamma_r (c)$ satisfies the linear second-order  differential equation  ( $ r \geq 0 $) 
\be \frac{d^2 \hat \Gamma_r(c)}{d c^2} +\frac{c}{24}\frac{d \hat \Gamma_r(c)}{d c}-  \frac{r}{24} \hat \Gamma_r (c)=0. \label{che} \ee
The orthogonal polynomial solutions (\ref{rec}) of (\ref{he}) have no real root for even degree and zero is the only real root for odd degree. We list a few polynomials for reference. For example, 
\bean \hat \Gamma_0(c) &=&1, \quad  \hat \Gamma_1(c)=c,   \quad \hat \Gamma_2(c)=(c^2+24), \\
 \hat \Gamma_3(c)&=& c^3+72c, \quad \hat \Gamma_4(c)=c^4+144 c^2+1728, \quad \hat \Gamma_5(c)=c^5+240 c^3+8640c,  \\
\hat \Gamma_6(c) &=& c^6+360 c^4+25920c^2+207360,  \quad \hat \Gamma_7(c)=c^7+504c^5+60480c^3+1451520c, \\
\hat \Gamma_8(c)& =& c^8+672 c^6+120960c^4+5806080c^2+34836480. \eean

\item For $ t \to \infty$, we consider $c=0$ in (\ref{pol}). Then $x_1=2w i  \sqrt{t}$ and $x_2=-12t+ 2wi \sqrt{t}$. Plugging (\ref{pol}) into $H_r$, we assume 
\be H_r(t)= \frac{\hat G_r (w)}{ r !} (2i)^{r} (t)^{r/2}+ O (\vert t \vert ^{\frac{(r-1)}{2}}), \label{ga4} \ee
where $ \hat G_r(w)$ is a polynomial in $w$. To find the polynomials $\hat G_r(w)$, in a similar way as (\ref{cca}) and (\ref{cca2}), we have 
\bea  w \hat G_r (w) &=&   \hat G_{r+1}(w)-6 r \hat G_{r-1}(w), \quad \hat G_0=1, \quad \hat G_1=w, \label{rec4}  \\
 \frac{d \hat G_r(w)}{d w} &= & r \hat G_{r-1}(w).  \label{rec5} \eea
When compared with the recursive relation (\ref{rec}), the equation (\ref{rec4}) has a minus sign before 6. By (\ref{rec4}) and (\ref{rec5}), it can be also seen that $\hat G_r (w)$ satisfies the linear second-order  differential equation  ( $ r \geq 0 $) 
\be \frac{d^2 \hat G_r(c)}{d c^2} + \frac{c}{6}\frac{d \hat G_r(c)}{d c}-  \frac{r}{6} \hat G_r (c)=0. \label{he3} \ee
The orthogonal polynomial solutions (\ref{rec4}) of (\ref{he3}) also have no real root for even degree and zero is the only real root for odd degree.  We list a few polynomials for reference. For example, 
\bean {\hat G}_0(w) &=&1, \quad  {\hat G}_1(w)=w,   \quad {\hat G}_2(w)=(w^2+6), \\
 {\hat G}_3(w)&=& w^3+18w, \quad {\hat G}_4(w)=w^4+36 w^2+108, \quad {\hat G}_5(w)=w^5+60 w^3+540w,  \\
{\hat G}_6(w) &=& w^6+90 w^4+1620w^2+3240,  \quad {\hat G}_7(w)=w^7+126 w^5+3780w^3+22680w, \\
{\hat G}_8(w)& =&w^8+168 w^6+7560w^4+90720w^2+136080.    \eean
\end{itemize}
Also, the  number of  real roots of the  Wronskian of the orthogonal polynomials  has been investigated \cite{gg, kz}. Let the multi-index $(n_1, n_2, n_3, \cdots, n_m )$ be related to the partition $\zeta= (\zeta_1, \zeta_2, \cdots  \zeta_m)$ by
\[ n_j= \zeta_j+ j-1, \quad j=1,2,3, \cdots, m,  \quad d_{\zeta}=p-q, \]  where $p, q$ are the numbers of odd and even elements in $n_j$,  respectively. One defines the length of the partition $\zeta$ is $\vert \zeta \vert =\sum_{j=1}^m \zeta_j= \zeta_1+\zeta_2+\cdots+ \zeta_m.$ For a symmetric case, the Wronskian $\phi_{\zeta}$ has the well-defined parity 
\be \phi_{\zeta}(-x)=(-1)^{\vert \zeta \vert} \phi_{\zeta}(x), \label{ron} \ee
where $x=c$ or $x=w$. For the symmetric property (\ref{ron}), we have the \\
{\bf Theorem} \cite{gg}: \\
(1) The number of simple real roots of the Wronskian is $\sum_{j=1}^m (-1)^{m-j}\zeta_j- \frac{\vert d_{\zeta} + (m-2 [m/2]) \vert }{2} $.\no \\ 
(2) The multiplicity of 0 is $  \frac{d_{\zeta} (d_{\zeta}+1)}{2}. $ \\
For $m=1$ and $n_1=n$, one has the cases $\Gamma_{r} (w)$ and $G_{r}(c)$. It is not difficult to see that (\ref{tw}) also satisfies the symmetric property (\ref{ron}) as $ t \to \pm \infty$ for $\Gamma_{r} (w)$ and $G_{r}(c)$. 
\begin{itemize}
\item $n=2k+1 (k \geq 2):$ We have $ d_{\zeta}=2 $ and $ (\zeta_1, \zeta_2)=(1,2k)$. From the {\bf Theorem}, one knows that there are $2k-1-1=2k-2$ simple real roots, and $(0,0)$ is the triple roots. So we obtain $2k-2+3=2k+1$ real roots as $ t \to -\infty$ for $\Gamma_{r} (w)$ and $ t \to \infty$  for  $G_{r}(c)$. For example, taking $n=5$, we have as $ t \to \infty$ 
\[\Omega_{(1,5)}(c,t) = det 
 \left(\ba{cc} ct+o(t)  & \frac{G_5(c)}{5!} t^{5/2}+ o(t^{5/2}) \\
                 1 & \frac{G_4(c)}{4!} t^{2} +o(t^2) \ea \right)=( \frac{1}{30}c^5-4c^3)  t^{5/2} + o(t^{5/2}). \]
The equation $ \frac{1}{30}c^5-4c^3=0$ has five real root, and $c=0$ is a triple root.
\item $n=2k (k \geq 1):$ We have $ d_{\zeta}=0 $ and $ (\zeta_1, \zeta_2)=(1,2k-1)$. From the {\bf Theorem}, one knows that there are $2k-1-1-0=2k-2$ simple real roots, and $(0,0)$ is not a root. In fact, another two real roots can be obtained from $\hat \Gamma_{r} (c)$ and $ \hat G_{r}(w)$ as $ t \to \pm \infty$. It can be seen as follows. For example, one considers $t \to \infty $. Using (\ref{tw}), (\ref{rec5}) and (\ref{ga4}), we know that
\bean \Omega_{(1,2k)}(x,y,t) &=& 2wit^{1/2} \frac{\hat G_{2k-1}(w)}{ (2k-1) !} (2i)^{2k-1} t^{(2k-1)/2}-\frac{\hat G_{2k}(w)}{ (2k) !} (2i)^{2k} t^{k} + O (\vert t \vert ^{\frac{(2k-1)}{2}}) \\
&=& t^k (2i)^{2k}\frac{1}{(2k)!} ( 2kw \hat G_{2k-1}(w)-\hat G_{2k}(w)) + O (\vert t \vert ^{\frac{(2k-1)}{2}})              \\
&=& t^k (2i)^{2k}\frac{1}{(2k)!} ( w \frac{d\hat G_{2k}(w)}{dw}-\hat G_{2k}(w))+ O (\vert t \vert ^{\frac{(2k-1)}{2}})
\eean
From (\ref{rec4}), it is known that  
\[\hat G_{2k}(w)=\alpha_{2k}w^{2k}+\alpha_{2(k-1)}w^{2(k-1)}+ \alpha_{2(k-2)}w^{2(k-2)}+ \cdots+ \alpha_2 w^2+ \alpha_0, \]
where $\alpha_{2k}, \alpha_{2(k-1)}, \cdots, \alpha_2, \alpha_0$ are positive integers. Therefore, 
\be w \frac{d\hat G_{2k}(w)}{dw}-\hat G_{2k}(w) =L(w)-\alpha_0, \label{dea} \ee
where
\[ L(w)= (2k-1)\alpha_{2k}w^{2k}+(2k-3)\alpha_{2(k-1)}w^{2(k-1)}+(2k-5) \alpha_{2(k-2)}w^{2(k-2)} +\alpha_2 w^2. \]
Since $(2k-1)\alpha_{2k}, (2k-3)\alpha_{2(k-1)}, (2k-5) \alpha_{2(k-2)}, \cdots, \alpha_2$  are positive integers, we obtain $ L(w) \geq 0 $, and  the only minima is at $w=0$. By $\alpha_0 > 0$, it is known that the equation (\ref{dea}) only has two real roots. As $ t \to -\infty$, one can consider similarly for $ \hat \Gamma_r(c)$. 
\end{itemize}

One remarks here that if $ n_1 \neq 1 $, then some real roots of (\ref{com}) come from neither $c=0$ nor $w=0$. For example, we take $n_1=2, n_2=4$ as an example. Then $\Omega_{(2,4)}(x,y,t)=H_2H_3-H_1H_4$ and $q=1/2$. When one considers $t \to \infty$ and uses (\ref{pol}), one has 
\bean M(c,w) &=& 1/24\,{c}^{5}-2\,{c}^{3}-5/3\,{c}^{3}{w}^{2}+72\,c+10/3\,c{w}^{4}+24\,
c{w}^{2}=0,  \\
 R(c,w)  &=&  {\frac {5}{12}}\,{c}^{4}w-10/3\,{w}^{3}{c}^{2}-12\,{
c}^{2}w+144\,w+16\,{w}^{3}+4/3\,{w}^{5}=0.\eean
They have $\rho=2+4-1=5$ real roots, and some real roots $ (c,w) \neq (0,0)$. For $t \to -\infty$, it can be considered similarly, and also some real roots $ (c,w) \neq (0,0)$. \\

\section{Concluding Remarks}					
We study the asymptotic behaviors of lump solutions of KP-(I) equation using the Grammian determinant structure. In particular, for $m=1$ and $p=1$, it is found that the positions $\zeta$ of peaks can be described by the real roots of orthogonal polynomials; moreover, the peaks are on one  vertical line when time approaches - $\infty$, and then they will be on a horizontal line when time approaches $\infty$, i.e., there is a rotation $\pi/2$ after interaction.	Also, for $m=2$ and $p=1$, in some special case,  as $\vert t \vert \to \infty $,  the locations of peaks will depend on the real roots of Wronskian of the orthogonal polynomials. The asymptotic structures are different, depending on the partition structure (even or odd); however, there is also  a rotation $\pi/2$ after interaction. \\

\indent In \cite{ab3}, the lump solutions (rationally decaying potentials) of KP-(I) equation are constructed in a similar way. These solutions are classified by the pole structure of the corresponding
 meromorphic eigenfunctions and a set of integers of quantity called the charge. When comparing with their lump solutions, one has different function (\ref{fu}) and different recursive relation (\ref{gen}). Especially, the relation (\ref{eq}) is  important to get the orthogonal polynomials for the asymptotic analysis. There is no such relation in \cite{ab3}. To the best of my knowledge, this kind of orthogonal  polynomials have never been used in the asymptotic analysis for any rationally decaying solutions in  KP-(I) equation. \\

\indent In \cite{du1, du2, pi, pk}, the multi-rogue wave solutions of KP-(I) equation are constructed from the NLS hierarchy via different approach. It is the so-called NLS-KP correspondence. The authors choose appropriate parameters such that at some  intermediate time all the peaks will collide together to form the only extreme rogue wave, i.e., $P_m$ breather of the KP-(I) equation. On the other hand, by a similar consideration in \cite{oy2}, we also could choose $p_{\alpha}=q_{\alpha}=1, \alpha=1, 2, 3, \dots, m$ and $(n_1, n_2, n_3, \cdots, n_m )=(1, 3, 5, 7, \cdots, 2m-1)$ for some particular  parameters 	$c_{lk}$ to obtain the $P_m$ breather of the KP-(I) equation. As mentioned above, the asymptotic structures will be related to the partition structures. In \cite{pi2}, there are concentric rings with in the center Peregrine breathers for the NLS equation using different partitions and parameters. From this point of view, our asymptotic structures only depend on the partitions $\zeta$, that is, the exponent $q$ in (\ref{qd}) and the real roots of the polynomials $ M(c,w)$ and $ R(c,w) $ in (\ref{rea}). It could be interesting to investigate such structures in KP-(I) equation when compared with the NLS equation. The research in this direction will be published elsewhere.

\subsection*{Acknowledgments}
The author thanks for the valuable suggestions from Prof.V. B. Matveev, Prof. Y. Ohta and Prof. J. Yang.  This work is supported by the Ministry of  Science and Technology  under Grant No. 104-2115-M-606-001.

\end{document}